 \documentclass[final,5p,times,twocolumn,authoryear]{elsarticle}

\usepackage{amsmath,amssymb,amsfonts,xcolor}
\usepackage{algorithmic}
\usepackage{graphicx}
\usepackage{textcomp}
\usepackage{tikz}     % Core TikZ package
\usetikzlibrary{arrows,shapes,positioning,calc,backgrounds,decorations.markings,quotes,angles}

\newcommand{\bu}{\mathbf{u}}

\newtheorem{ass}{\bf Assumption}
\newtheorem{prop}{\bf Proposition}
\newtheorem{cor}{\bf Corollary}
\newtheorem{lem}{\bf Lemma}

\newcommand{\e}{\ \vskip 1.5mm \noindent}

\usepackage{amssymb,amsmath}
\usepackage{lipsum}
\journal{arXiv}

\begin{document}

\begin{frontmatter}



\title{On relaxing the $N$-Reachability Implicit Requirement in NMPC Design}


\author[first]{Mazen Alamir}
\affiliation[first]{organization={Univ. Grenoble Alpes, CNRS, Grenoble INP, GIPSA-lab, 38000 Grenoble, France},%Department and Organization
            addressline={}, 
            city={Grenoble},
            postcode={38000}, 
            state={},
            country={France}}

\begin{abstract}
This paper proposes a proof of stability for Model Predictive Control formulations involving a prediction horizon that might be too short to meet the reachability condition generally invoked as a sufficient condition for closed-loop stability. This condition is replaced by a contraction condition on the stage cost. But unlike the contraction based existing formulations where the prediction horizon becomes a decision variable, the formulation proposed in this paper remains standard in that it uses constant and short prediction horizon. An illustrative example is provided to assess the relevance of the proposed formulation. 
\end{abstract}




\end{frontmatter}

\section{Introduction}
\label{sec:introduction}
The closed-loop stability of dynamical systems under Nonlinear Model Predictive Control (NMPC) is by now a mature topics as far as enumerating sufficient conditions is concerned. One of the most systematically invoked sufficient conditions for theoretical stability of NMPC schemes is the so-called \textit{$N$-reachability condition}\cite{Muller2014,kohler2018nonlinear,Mayne2000, rawlings2017model}. This condition stipulates that the \textit{targeted/desired} state, say $x^{\text{ref}}$, is reachable in less that $N$ steps from any state belonging to some subset $\mathbb X$ of initial states of interest. Technically, the $N$-reachability  assumption is used in the stability proof to infer the existence, for any initial state $x\in \mathbb X$, of a sequence of input actions $\tilde u(x):=(u_0, \dots, u_{N-1})$ that steers the state from $x$ to the desired value $x^{\text{ref}}$. The cost value associated to $\tilde{u}(x)$ hence serves as an upper bound on the optimal cost at the next state and, from this, the stability proof generally follows thanks to the Lassalle's principle \cite{la1966invariance} through some well known sequence of technical arguments.
\e 
Obviously, when saturation constraints are present, the satisfaction of this reachability assumption depends on the distance from $x$ to the desired state $x^{\text{ref}}$. This observation is not new and as far as theoretical developments are concerned, it triggered research works on the concept of \textit{reference/command-governor}  \cite{garone2017reference, klauvco2017real} in the MPC literature. In a nutshell, this  amounts at amending the user-defined set-point so that a constraints-compatible modified set-point is computed and sent to the original NMPC controller. Traditionally the latter solution induces an important extra-computational cost due to the new (or extended) optimization problem to solve. This makes the additional scheme eligible to be one of those theoretical components that are generally discarded in real-life implementations leaving to the final user the duty of feeding the NMPC controller with \textit{reasonably filtered} set-points. 
\e 
This paper proposes an NMPC formulation involving almost no additional computational cost that smoothly enforces the system to \textit{get closer} to some reachable steady state that is not explicitly computed and which is smoothly and automatically updated as time goes and as a natural consequence of the formulation so that the system asymptotically reaches a neighbourhood of the originally targeted state. This is done mainly by adding  a specific terminal cost that incorporates a penalty on the derivative of the terminal state.
\e The idea of penalizing the final derivative's amplitude has been recently proposed  \cite{alamir2021new} in the context of economic MPC,  although in a different  framework and using different set of sufficient conditions. In particular, the $N$-reachability condition is required in \cite{alamir2021new} to derive closed-loop stability while in the present contribution, a new set of intuitive sufficient conditions is proposed that enable to replace the $N$-reachability assumption for all pairs $(x_0,x^{\text{ref}})\in  \mathbb X^2$ by the $N$ reachability, from $x_0$ of at least, one steady state with lower stage cost. This suggestion provides a simple scalar weight that can be added to the set of setting parameters of MPC that can be then tuned using the certification scheme such as the one proposed in \cite{alamir2024derivation}. In spite of the fact that the proposed formulation incorporates a contraction condition on the stage cost, it strongly differs from the existing proposed contraction-based NMPC \cite{alamir2017contraction,polver2025robust} in that it does not consider the prediction horizon as a part of the decision variable to be updated at each decision instant. Moreover, it does not require a specific contraction-related storage function to be chosen as the contraction is related to the original stage cost. 
\e
{  As in almost all provably stable formulations, some of the assumptions are difficult to check theoretically. This makes a last statistical certification step such as the one proposed in \cite{alamir2024derivation} mandatory to any NMPC formulation. This is the only way to get a solid assessment of the stability,  the real-time implementability as well as the constraint satisfaction. The theoretical foundations provided in this paper give rationale to consider this formulation as a candidate for such a certification procedure.}
% Beside the above mentioned contribution, the formulation adopted in the present paper explicitly acknowledges the fact that during the computation time $[k,k+1)$, the control $u_k$ cannot be viewed as a decision variable. Instead, it should be considered as a part of an extended state $z_k:=(x_k,u_k)$ that cannot be modified making the control $u_{k+1}$ to be applied at the next sampling interval $[k+1,k+2]$, the first truly freely available degree of freedom within the future sequence of input actions. This leads to a rather non conventional notation that strongly differs from standard texts on provably stable NMPC formulations. 
\e  
The paper is organized as follows: In Section \ref{sec-notation}, some definitions and notation are given and the problem is stated. Section \ref{sec-workingass} states and discusses the working assumption. Section \ref{secmainresults} provides the proof of the main results. Section \ref{secillustrative} provides an illustrative use-case assessing the relevance of the proposed ideas. Finally, the paper ends with a conclusion that summarizes the paper's finding. 
\section{Definitions and Notation}\label{sec-notation}
\begin{figure}[h]
\begin{center}
\begin{tikzpicture}[scale=1]
\draw[->, >=stealth, thick](-0.2,0) -- node[pos=0.95, below]{time}(5,0);
\draw[thick] (0,-0.2) node[below]{\footnotesize $k$} -- (0,0.2);
\draw[thick] (1.3,-0.2) node[below]{\footnotesize$k+1$} -- (1.3,0.2);
\draw[thick] (2.6,-0.2) node[below]{\footnotesize$k+1$} -- (2.6,0.2);
\draw[thick] (3.9,-0.2) node[below]{\footnotesize$k+2$}-- (3.9,0.2);
%-----
\draw[thin] (1.3,0) -- (1.3,1.5);
\draw[-] (0,0) -- (0,1.5);
\draw[-] (2.6,0) -- (2.6,1.5);
\draw[fill=black] (0,1) circle (1pt) node[left]{$x_k$};
\draw[fill=black] (1.3,1.35) circle (1pt) node[left]{$x_{k+1}$};
\draw[thin] (0,0.5) -- +(1.3,0);
\node[left] at(0,0.5) {$u_k$};
\draw[very thick] (1.3,0.65) -- node[above,pos=0.5]{\footnotesize $K(x_k,u_k)$} +(1.3,0);
%-----
\draw[<->] (2.6,0.3) -- node[above]{$\tau$}(3.9,0.3);
\draw[<->] (0,-0.6) -- node[below]{\footnotesize Solve $P(x_k,u_k)$}(1.3,-0.6);
\draw[->, dashed, thick] (1.35,-0.8) -- ++(0.65,0) -- ++(0,1.35);
\end{tikzpicture}
\end{center}
\caption{Schematic of  real-time compatible definition of MPC which amounts at computing the control to be applied at the expected state $x_{k+1}$ given the current state $x_k$ and the currently applied control $u_k$.}\label{nonconventional}
\end{figure}
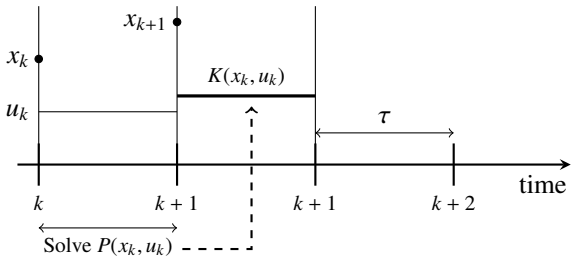

We consider a nonlinear discrete-time system of the form:
\begin{equation}
x_{k+1}=f(x_k,u_k) \quad (x_k,u_k)\in \mathbb R^{n}\times \mathbb R^{m}\label{dtsyst}
\end{equation}
that is supposed to be obtained through time discretization of a continuous time Ordinary Differential Equations (ODEs) of the form $\dot x=f_c(x,u)$. 
\e Standard MPC formulations, devoted to closed-loop stability analysis, consider that the control $u_k$, to be applied over the sampling interval $[k,k+1]$, is the outcome of the solution of an optimal control problem $P(x_k)$ that is parameterized by the current state $x_k$. This ideal setting does not acknowledge the fact that this solution step involves non vanishing computation time that might span the whole time period $[k,k+1]$ making a computed $u_k$ not available until the end of the sampling period.
\e 
That is the reason why, in this contribution, we consider the more realistic real-time compatible formulation depicted in Fig. \ref{nonconventional} where the control $u_k$ is supposed to be given from a previous computation step to be applied over $[k,k+1]$ during which an optimal control problem $P(x_k,u_k)$ is solved which is parameterized by the pair $(x_k,u_k)$ that encapsulates the \textit{given parameters} of the problem at instant $k$. 
\e 
It comes out from the previous discussion that a more realistic formulation of MPC should involve the following extended internal state:
\begin{equation}
z:= \begin{bmatrix} 
x\cr u
\end{bmatrix}\in \mathbb R^{n+m}  \label{defdez}
\end{equation}
that is governed by the following dynamics\footnote{Based on \eqref{defdez}, the notation $f(z)$ and $f_c(z)$ to shortly designate $f(x,u)$ and $f_c(x,u)$ respectively.}:
\begin{equation}
z_{k+1}=F(z_k,u_{k+1}):= \begin{bmatrix} 
f(z_k)\cr u_{k+1}
\end{bmatrix}  \label{defdeFz}
\end{equation}
In this setting, the degrees of freedom involved in any optimal control problem over a prediction horizon of length $N$, formulated at instant $k$, is given by:
\begin{equation}
\bu_{k}:=(u_{1\vert k},\dots,u_{N\vert k}) \in \mathbb R^{Nm}\label{defdeutildebf}
\end{equation}
where $u_{i\vert k}$ is planned to be applied over the time interval $[k+i, k+i+1]$.
\e In the sequel, the notation $n_z=n+m$ is used to denote the dimension of the extended state $z$.
\e Having such a sequence $\bu:=(u_1,\dots,u_N)\in \mathbb R^{Nm}$ together with some initial extended state $z$, the notation $z^\bu_i(z)$ is defined as follows:
\begin{equation}
z^\bu_0(z)=z\quad;\quad  z^\bu_{i+1}(z)=F(z^\bu_{i}(z), u_{i+1})
\end{equation}
for $i\in \{0,\dots,N\}$.
% \e In the remainder of the paper, the following notation is used:
% \begin{equation}
% \forall z=\begin{bmatrix} 
% x\cr u
% \end{bmatrix}:\qquad  z^+(z):=\begin{bmatrix}
% F(z)\cr u
% \end{bmatrix}\label{defdez+}
% \end{equation}
% which represents the extended state that is reached when applying the same control over the next prediction horizon.
% \begin{figure}[h]
% \begin{center}
% \begin{tikzpicture}[scale=1.4]
% \draw[->, >=stealth, thick](-0.2,0) -- node[pos=0.95, below]{time}(4,0);
% % \draw[thick] (0,-0.2) node[below]{\footnotesize $k$} -- (0,0.2);
% % \draw[thick] (1.3,-0.2) node[below]{\footnotesize$k+1$} -- (1.3,0.2);
% % \draw[thick] (2.6,-0.2) node[below]{\footnotesize$k+1$} -- (2.6,0.2);
% % \draw[thick] (3.9,-0.2) node[below]{\footnotesize$k+2$}-- (3.9,0.2);
% %-----
% \draw[thin] (1.3,0) -- (1.3,1.5);
% \draw[-] (0,0) -- (0,1.5);
% \draw[-] (2.6,0) -- (2.6,1.5);
% \draw[fill=black] (0,1) circle (1pt) node[left]{$x$};
% \draw[fill=black] (1.3,1.35) circle (1pt) node[left]{\scriptsize $x^+=F(z)$};
% \draw[thin] (0,0.5) -- +(1.3,0);
% \node[left] at(0,0.5) {$u$};
% \draw[thick] (1.3,0.5) -- node[above,pos=-0.1]{$u$} +(1.3,0);
% %-----
% \draw[<->] (-0.3,1.1) -- node[left]{$z$}(-0.3, 0.4);
% \draw[<->] (1.4,1.45) -- node[right]{$z^+(z)$}(1.4, 0.35);
% \end{tikzpicture}
% \end{center}
% \caption{Definition of $z^+(z)$ given by \eqref{defdez+}.}\label{nonconventional_bis}
% \end{figure}
\e
In this paper, we consider a cost function of the form:
\e \hrule \e
\begin{align}
&J(\bu\ \vert\   z):= \Phi_\gamma\bigl(z^{\bu}_N(z)\bigr)+\underbrace{\sum_{i=0}^N \ell(z^\bu_i(z))}_{ S^\bu(z)}\label{defdeJ} \\
&\text{where}\quad \Phi_\gamma(\xi):= \gamma^2\|f_c(\xi)\|^2 + \gamma \ell(\xi)\label{defdePhi}
\end{align}
\e \hrule \e
This cost function is used to define the following optimal control problem that is parameterized by the initial extended state $z$:
\begin{equation}
\text{P}(z):\quad \bu^\star(z)\leftarrow  \text{arg}\min_{\bu\in \mathbb U^N} J(\bu\ \vert \ z)\label{defdePz}
\end{equation}
where $\mathbb U\subset \mathbb R^{m}$ is some compact set of admissible control inputs. The optimal value of $J$ at $z$ is denoted by $J^\star(z)$. Notice that in the formulation of this paper, the state constraints are supposed to be transformed into soft constraints which are represented in the cost function via exact penalty terms \textit{inside} the expression of $\ell(z)$. 
\e Once P$(z_k)$ is solved during $[k,k+1)$ to yield the optimal sequence denoted by: $$\bu^\star(z)=(u^\star_1(z_k),\dots, u^\star_N(z))$$
the MPC feedback amounts at applying  $u^\star_1(z_k)$ over the sampling interval $[k+1,k+2)$. In the sequel, the following short notation is used:
\begin{equation}
z^\star_i(z)=z^{\bu^\star(z)}_i(z) \label{shortnotation}
\end{equation}
to designate the state, $i$-steps later, on the open-loop optimal trajectory starting at $z$. 
\e 
\textsc{Further notation}.
\e 
The following parameterization of compact subsets in the extended state space is used in the sequel:
\begin{equation}
\mathbb Z^\star_\eta:= \Bigl\{z\in \mathbb R^{n+m}\ \vert\  J^\star(z)\le \eta\Bigr\} \label{defdeZeta}
\end{equation}
\e  Given a compact set $\mathbb Z$ in the extended state space, the following sequence of compact sets can be defined recursively by $\mathbb Z^{(+i)}:=\Bigl\{F(z)\ \vert\ z\in \mathbb Z^{(+(i-1))}\Bigr\}$ with the initialization $\mathbb Z^{(+0)}=\mathbb Z$. 
Notice that $\mathbb Z^{(+i)}$ represents the set containing all possible values of the extended state $i$-step ahead starting from $\mathbb Z$ under a sequence of $i$ control inputs contained in $\mathbb U$. The union of all these sets up to $i=N$ is simply denoted hereafter by $\mathcal Z(\mathbb Z)$, namely: $\mathcal Z(\mathbb Z):=\bigcup_{i=0}^N \mathbb Z^{(+i)}$
which is a compact set that contains all possible extended states resulting from an initial state in $\mathbb Z$ after a number of steps lower than or equal to $N$.
\section{Working assumptions}\label{sec-workingass}
The first assumption concerns the regularity of the maps being involved. It is necessary for the existence of solutions to the optimization problems underlying the NMPC formulation given by \eqref{defdeJ}-\eqref{defdePz} as well as to define bounds on some terms appearing here and there in the proof. \\
\begin{ass}\label{assreg}
The maps $f_c$, $f$ and $\ell$ are Lipschitz over any compact set. More precisely, for all $g\in \{f_c, f, F, \ell\}$ the following inequality:
\begin{equation}
\|g(z_1)-g(z_2)\| \le K_{g}(\mathbb Z)\times \|z_1-z_2\|\label{lypchitz}
\end{equation}
holds true for all $(z_1,z_2)\in \mathbb Z^2$. $\hfill \heartsuit$
\end{ass}
\ \\
The next assumption concerns the  stage cost:
\begin{ass}\label{asscompactness}
The stage cost is positive, unbounded at infinity and is such that $\ell(z)=0$ implies that $z$ is a steady pair, namely $f_c(z)=0$.$\hfill \heartsuit$
\end{ass}
\ \\
The next assumption is the major \textit{relaxed} reachability assumption:
\begin{ass}\label{asscontraction}
There exists a tuple: $$(J_0, N, \alpha)\in \mathbb R_+\times \mathbb N\times (0,1)$$ such that $\mathbb Z^\star_{J_0}\neq \emptyset$  and for all $z\in \mathbb Z^\star_{J_0}$, there exists $\tilde\bu(z)\in \mathbb U^N$ satisfying the following conditions:\\
\begin{enumerate}
    \item The terminal state is stationary, namely: 
    \begin{equation}
f_c(z^{\tilde\bu(z)}_N)=0 \label{condfinalfcx}
\end{equation}
    \item The terminal stage cost satisfies:
    \begin{equation}
\ell(z^{\tilde\bu(z)}_N(z))\le \alpha \ell(z) \label{condcontraction}
\end{equation}
\end{enumerate}
when multiple candidates of $\tilde\bu$ exist that satisfy \eqref{condfinalfcx} and \eqref{condcontraction}, the sequence implicitly referred to by $\tilde\bu(z)$ is the one that minimizes the term $S^{\tilde\bu}(z)$.$\hfill \heartsuit$
\end{ass}
\ \\
Notice that the standard reachability assumption corresponds to the specific case where $\alpha=0$. Indeed, in this case, the targeted steady state satisfying $f_c(z^d)=0$ and $\ell(z^d)=0$ would be $N$-reachable under the dedicated sequence $\tilde\bu(z)$.
% \e Assumption \ref{asscontraction} enables the following definition to be stated:
% \e 
% \begin{definition}
% For all $z\in Z_{J_0}$, the following notation are used in the sequel:
% \begin{equation}
% J_0:=\max_{z\in \mathbb Z^\star_{J_0}} \Bigl[J(\tilde\bu(z)\ \vert\ z)\Bigr] \label{defdeeta0}
% \end{equation}
% \end{definition}
\e The next assumption is related to the existence of a bound on the cumulated stage cost term $S^{\tilde\bu(z)}(z)$ associated to an initial extended state $z\in \mathbb Z^\star_{J_0}$ under the sequence $\tilde\bu(z)$ invoked in  Assumption \ref{asscontraction}. Regarding this cost, the following implication obviously holds:
\begin{equation}
\Bigl\{\ell(z)=0\Bigr\}\quad \Rightarrow\quad S^{\tilde\bu(z)}(z)=0 \label{implication}
\end{equation}
since in this case, $z=(x,u)$ is a steady pair and hence the choice $\tilde\bu:=(u,u,\dots,u)$ leads to zero cost and terminal state $z$ that meets the requirements \eqref{condfinalfcx} and \eqref{condcontraction}.
\e The following assumption is simply a specific instance of the above implication expressed by \eqref{implication}:
\ \\
\begin{ass}\label{assimplication}
There exists positive $\kappa_1$ such that for all $z\in \mathbb Z^\star_{J_0}$, the following inequality:
\begin{equation}
S^{\tilde\bu(z)}(z)\le \kappa_1\ell(z)\label{lkjhlkjh76}
\end{equation}
holds  true. $\hfill \heartsuit$
\end{ass}
\ \\
The last Assumption we need is rather technical although intuitively sound. It can be \textit{informally} stated as follows: If a sequence $\tilde\bu\in \mathbb U^N$ leads to a sufficiently small neighbourhood of a steady state, then there is a \textit{small} deformation of $\tilde\bu$ that leads to a rigorously steady state with close stage cost. More formally:
\ \\
\begin{ass}\label{assimplicit}
There is sufficiently small $\varepsilon_0>0$ such that, for all $(z,\tilde\bu)\in \mathbb Z^\star_{J_0}\times \mathbb U^N$  satisfying:
\begin{equation}
\|f_c(z^{\tilde\bu}_N(z))\|\le \varepsilon\le \varepsilon_0 \label{ineqvarpesilon}
\end{equation}
there exists a feasible sequence of control $\check\bu$ satisfying:
\begin{equation}
\|f_c(z^{\check\bu}_N(z)\|=0\ \text{and}\ \ell(z^{\check\bu}_N(z))\le \ell(z^{\tilde\bu}_N(z))+O(\varepsilon) \label{inversion}
\end{equation}
\end{ass}
\e This ends the presentation of the working assumptions that are needed to derive the main results which can be summarized as follows: 
\begin{itemize}
\item Assumption \ref{assreg} is a rather standard regularity assumption.
\item Assumption \ref{asscompactness} is satisfied by all regulation-like stage costs. 
\item Assumption \ref{asscontraction} is strictly less stringent than the standard $N$-reachability of $\ell=0$ that holds in the specific case where $\alpha=0$.
\item {  Assumption \ref{assimplication} is a specific instantiation of the implication \eqref{implication} that is a consequence of the previous assumptions \ref{assreg}-\ref{asscontraction}. The linear form of the r.h.s of \eqref{lkjhlkjh76} encompasses a large set of possibilities for which it is possible to upper bounding the non-linear dependency by the linear form for sufficiently high $\kappa_1$.}
\item Finally Assumption \ref{assimplicit} is a reformulation of the implicit function theorem in a non necessarily differentiable setting. A variant of this assumption has already been discussed in \cite{alamir2021new}. {  This assumption might not be satisfied for a class of systems that are not \textit{regular enough}.}
\end{itemize}
\section{Main results}\label{secmainresults}
The chain of arguments used in the proof is standard. Namely, the first result uses the property of the trajectory resulting from the sequence of inputs $\tilde\bu$ invoked in Assumption \ref{asscontraction} in order to characterize some terminal properties of the optimal solution. Then this characterization is used in order to prove that the optimal solution is such that the optimal cost decreases along the closed-loop trajectory with a rate that is proportional to $\ell(z)$.
\begin{center}
\begin{tikzpicture}
\node[rounded corners, fill=black!5, inner sep=3mm]{
\begin{minipage}{0.45\textwidth}
\begin{lem}\label{lemcharactilde}
For all $z\in \mathbb Z^\star_{J_0}$, the optimal solution to $P(z)$ satisfies the following terminal conditions:
\begin{subequations}
\begin{align}
\|f_c(z^\star_N(z)\|^2&\le \dfrac{1}{\gamma^2}\Bigl[\gamma\alpha+\kappa_1\Bigr]\times \ell(z)\label{ineqfc}\\
\ell(z^\star_N(z))&\le \Bigl[\alpha+\dfrac{\kappa_1}{\gamma}\Bigr]\times \ell(z)\label{ineqfc2}
\end{align}
{  which also leads to 
\begin{equation}
\|f_c(z^\star_N(z))\|\le \dfrac{1}{\sqrt{\gamma}}\ell^{1/2}(z^\star_N(z)) \label{bonnerelation}
\end{equation}}
\end{subequations}
\end{lem}
\end{minipage}
};
\end{tikzpicture}
\end{center}
\e \text{\sc Proof}. Since $z\in \mathbb Z^\star_{J_0}$, Assumption \ref{asscontraction} implies that there exists $\tilde\bu(z)$ satisfying \eqref{condfinalfcx} and \eqref{condcontraction}. Therefore, given the definition \eqref{defdeJ} of the cost function, it comes that:
\begin{equation}
J(\tilde\bu(z)\ \vert\  z)= \underbrace{\Phi_\gamma(z^{\tilde\bu}_N(z))}_{\le 0+\gamma\alpha\ell(z)}+S^{\tilde\bu(z)}(z) \label{lem11}
\end{equation}
This together with \eqref{lkjhlkjh76} of Assumption \ref{assimplication} leads to:
\begin{equation}
J(\tilde\bu(z)\ \vert z)\le [\gamma\alpha+\kappa_1]\times \ell(z)\label{hgy76}
\end{equation}
and since the optimal cost is lower than $J(\tilde\bu(z)\ \vert z)$ and because $\ell$ is positive and hence so is $S^{\bu^\star(z)}(z)$, it comes out by \eqref{defdeJ} and \eqref{defdePhi}:
\begin{equation}
\gamma^2\|f_c(z^\star_N(z)\|^2+\gamma\ell(z^\star_N(z)) \le [\gamma\alpha+\kappa_1]\times \ell(z) \label{hg743}
\end{equation}
and since each of the terms at the l.h.s of \eqref{hg743} is positive, each term should be lower that the r.h.s. This obviously gives \eqref{ineqfc} and \eqref{ineqfc2}. $\hfill\Box$
\e Based on the previous lemma the following corollary can be proved that characterizes the next state $z_1^\star(z)$ on the closed-loop trajectory:
\e 
\begin{center}
\begin{tikzpicture}
\node[rounded corners, fill=black!5, inner sep=3mm]{
\begin{minipage}{0.45\textwidth}
\begin{cor}\label{lecoroll}
For all sufficiently small $\varepsilon>0$, there exists sufficiently high $\gamma$ such that for all $z$ in $\mathbb Z^\star_{J_0}$, the pair $(z^\star_1(z),\bu^+)$ in which:
\begin{equation}
\bu^+=(u^\star_2(z),\dots,u^\star_N(z),u^\star_N(z))\in \mathbb U^N \label{defdeubstarplus}
\end{equation}
satisfies \eqref{ineqvarpesilon} of Assumption \ref{assimplicit}.
\end{cor}
\end{minipage}
};
\end{tikzpicture}
\end{center}
\e \text{\sc Proof}. We need to prove that for sufficiently high $\gamma$, one gets:
\begin{equation}
\|f_c\bigl(z^{\bu^+}_{N}(z^\star_1(z))\bigr)\|\le \varepsilon \label{torpove}
\end{equation}
Indeed, since the first $N-1$ states and control inputs are those of the previous optimal trajectory at instant $2,\dots,N$, one has:
\begin{equation}
z^\star_{N-1}(z^\star_1(z))=z^\star_{N}(z) \label{evidence}
\end{equation}
Consequently, using Lemma \ref{lemcharactilde}, the following two inequalities stem directly from \eqref{ineqfc}-\eqref{ineqfc2} together with \eqref{evidence}:
\begin{align}
\|f_c\bigl(z^{\bu^+}_{N-1}(z^\star_1(z))\bigr)\|^2 \le \dfrac{1}{\gamma^2}\Bigl[\gamma\alpha+\kappa_1\Bigr]\times \ell(z)\label{jhbv420}\\
\ell\bigl(z^{\bu^+}_{N-1}(z^\star_1(z))\bigr)\le \Bigl[\alpha+\dfrac{\kappa_1}{\gamma}\Bigr]\times \ell(z)\label{mmm43}
\end{align}
Now using Assumption \ref{assreg}:
\begin{align}
&\|f_c\bigl(z^{\bu^+}_{N}(z^\star_1(z))\bigr)\|^2\le \|f_c\bigl(z^{\bu^+}_{N-1}(z^\star_1(z))\bigr)\|^2 +\nonumber \\ 
&\qquad\qquad\quad+\bigl[K_{\|f_c\|^2}(\mathcal Z(\mathbb Z^\star_{J_0}))\bigr]\times \|\Delta z^{\bu^+}_{N}(z^\star_1(z))\|\label{knbv72}
\end{align}
in which:
\begin{align}
\Delta z^{\bu^+}_{N}(z^\star_1(z))&:=z^{\bu^+}_{N}(z^\star_1(z))-z^{\bu^+}_{N-1}(z^\star_1(z))\nonumber \\
&=f(z^{\bu^+}_{N-1}(z^\star_1(z)))-z^{\bu^+}_{N-1}(z^\star_1(z))\\
&=O\bigl(\|f_c(z^{\bu^+}_{N-1}(z^\star_1(z)))\|\bigr)
\end{align}
On the other hand, thanks to \eqref{jhbv420} and since $\ell(z)$ is bounded over $\mathbb Z^\star_{J_0}$, it comes out that $\|\Delta z^{\bu^+}_{N}(z^\star_1(z))\|=O(1/\sqrt{\gamma})$. Using this in \eqref{knbv72} leads to:
\begin{align*}
\|f_c\bigl(z^{\bu^+}_{N-1}(z^\star_1(z))\bigr)\|^2&\le  \dfrac{1}{\gamma^2}\Bigl[\gamma\alpha+\kappa_1\Bigr]\times \max_{z\in \mathbb Z^\star_{J_0}}[\ell(z)] + \nonumber \\
&+\bigl[K_{\|f_c\|^2}(\mathcal Z(\mathbb Z^\star_{J_0}))\bigr]\times O(1/\sqrt{\gamma})
\end{align*}
Therefore, by taking $\gamma$ sufficiently high, it is possible to make $\|f_c\bigl(z^{\bu^+}_{N}(z^\star_1(z))\bigr)\|$ smaller than $\varepsilon$ which proves \eqref{torpove}. $\hfill\Box$
\e Having at hand the previous preliminary results, the following main result can be established:
\e 
\begin{center}
\begin{tikzpicture}
\node[rounded corners, fill=black!5, inner sep=3mm]{
\begin{minipage}{0.45\textwidth}
\begin{prop}
For sufficiently high $\gamma$, for all initial extended state $z_0\in \mathbb Z^\star_{J_0}$
the closed-loop trajectory $\{z^\text{cl}_k\}_{k=0}^\infty$ under the MPC feedback defined by \eqref{defdeJ}-\eqref{defdePz} is such that:
\begin{align}
&  J^\star(z^\text{cl}_{k+1})-J^\star(z^\text{cl}_{k})\lesssim -\beta \ell(z^\text{cl}_k)\ \text{as $\gamma\rightarrow\infty$}\label{decrease}\\
&  \lim_{k\rightarrow\infty}\ell(z^\text{cl}_k)\sim 0 \text{as $\gamma\rightarrow\infty$} \label{asymptotic}
\end{align}
where $\beta>0$.
\end{prop}
\end{minipage}
};
\end{tikzpicture}
\end{center}
{\sc Proof.} The proof proceeds by the following steps:
\begin{itemize}
\item First the inequality \eqref{decrease} is proved for $k=0$.
\item It is proved that the arguments can be repeated for $k>0$.
\item This proves \eqref{asymptotic}.
\end{itemize}
Notice that  \eqref{decrease} instantiated at $k=0$ concerns the comparison between the two terms $J^\star(z_0)$ and $J^\star(z^\star_1(z_0))$ as $z^\text{cl}_0=z_0$ and $z^\text{cl}_1=z^\star_1(z_0)$.
\e Let $\gamma_0>0$ be a sufficiently high number for the choice $\varepsilon_0=1/\sqrt{\gamma_0}\times (\max_{z\in \mathbb Z}[\ell(z)])$ to be sufficiently small in the sense of Assumption \ref{assimplicit}. Then  Corollary \ref{lecoroll} stipulates that there is a sufficiently higher $\gamma>\gamma_0$ making the sequence $\bu^+$ invoked in Corollary \ref{lecoroll} satisfying \eqref{ineqvarpesilon} of Assumption \ref{assimplicit} for {  $\varepsilon_0=\dfrac{1}{\sqrt{\gamma}}\ell^{1/2}(z^\star_N(z))$ by virtue of \eqref{bonnerelation}}. This sequence, when applied to $z^\star_1(z_0)$ leads to:
\begin{align}
f_c(z^{\check\bu}_N(z^\star_1(z_0)))&=0\label{ehoui} \\
\ell(z^{\check\bu}_N(z^\star_1(z_0)))&\le {  \underbrace{\ell\Bigl(\overbrace{z^\star_N(z_0)+O(1/\sqrt{\gamma}}^{z^{\bu^+}_N(z^\star_1(z_0)})\Bigr)+O(1/\sqrt{\gamma})}_{\sim \ell(z^\star_N(z_0))\  \text{as}\ \gamma\rightarrow \infty}}\label{jhu83}
\end{align}
\e 
Let us compute the cost $J(\check\bu\ \vert \ z^\star_1(z_0))$ of the candidate sequence $\check\bu$ when applied to $z^\star_1(z_0)$:
\begin{equation*}
J(\check\bu\ \vert\ z^\star_1(z_0))= \Phi_\gamma(z^{\check\bu}_N(z^\star_1(z_0)))+S^{\check\bu}(z^\star_1(z_0))
\end{equation*}
with:
\begin{align*}
\Phi_\gamma(z^{\check\bu}_N(z^\star_1(z_0)))&= \gamma^2\|\underbrace{f_c(z^{\check\bu}_N(z^\star_1(z_0)))}_{=0 \text{ [see  \eqref{ehoui}]}}\|^2+ \gamma \ell(z^{\check\bu}_N(z^\star_1(z_0)))\\
&{  \sim \gamma\ell(z^\star_N(z_0))}\quad \text{ \scriptsize [see \eqref{jhu83}]}\\
&{  \lesssim \Phi_\gamma(z^\star_N(z_0))-\gamma^2\|f_c(z^\star_N(z_0))\|^2}
\end{align*}
Therefore, the difference between the two terminal terms satisfies:
\begin{equation}
{  \Phi_\gamma(z^{\check\bu}_N(z^\star_1(z_0)))-\Phi_\gamma(z^\star_N(z_0))\lesssim 0\ \text{as $\gamma\rightarrow\infty$}}\label{deltaPhi}
\end{equation}
As for the cumulative stage cost, we have:
\begin{align}
S^{\check\bu}(z^\star_1(z_0))-&S^{\bu^\star(z_0)}(z_0)=\ell(z^{\check\bu}_N(z^\star_1(z_0)))-\ell(z_0) \nonumber \\
\text{ \scriptsize [Because of \eqref{jhu83}]}\qquad & \lesssim \ell(z^\star_N(z_0))-\ell(z_0)\ \text{as $\gamma \rightarrow\infty$}\label{paslast}
\end{align}
and using \eqref{ineqfc2}, we know that:
\begin{equation}
\ell(z^\star_N(z_0)) \le \Bigl[\alpha+\frac{\kappa_1}{\gamma}\Bigr]\times \ell(z_0) \label{hbgvgfc64}
\end{equation}
which, when injected in the \eqref{paslast} shows that:
\begin{equation*}
  S^{\check\bu}(z^\star_1(z_0))-S^{\bu^\star(z_0)}(z_0)\lesssim -(1-\alpha)\ell(z_0) \ \text{as $\gamma\rightarrow\infty$}
\end{equation*}
Therefore,
\begin{equation}
  S^{\check\bu}(z^\star_1(z_0))-S^{\bu^\star(z_0)}(z_0)\lesssim -\beta\ell(z_0)\ \text{as $\gamma\rightarrow\infty$}\label{enfinbeta}
\end{equation}
for some $\beta>0$.
\e Combining \eqref{deltaPhi} and \eqref{enfinbeta} together with the definition \eqref{defdeJ} of the cost function it comes that:
\begin{equation*}
  J(\check\bu\ \vert\ z^\star_1(z_0))\lesssim J^\star(z_0)-\beta\ell(z_0) \text{as $\gamma\rightarrow\infty$}
\end{equation*}
and since the optimal solution of $P(z^\star_1(z_0)$ is lower than this candidate value, it comes that:
\begin{equation*}
  J^\star(z^\star_1(z_0))\lesssim J^\star(z_0)-\beta\ell(z_0)\quad \text{  as $\gamma\rightarrow \infty$}
\end{equation*}
which proves \eqref{decrease} for $k=0$. 
\e Now in order to repeat the arguments for $k>0$, we need to prove that $z^\star_1(z_0)$ still belongs to $\mathbb Z^\star_{J_0}$ as long as $\ell(z_0)$ is not in a sufficiently small neighbourhood of $0$. But this is a direct consequence of the inequality: $$  J^\star(z^\star_1(z_0))\lesssim J^\star(z_0)-\beta\ell(z_0) \quad \text{  as $\gamma\rightarrow \infty$}$$ which implies that $J^\star(z^\star_1(z_0))$ remains lower than $J^\star(z_0)\le J_0$ (meaning that $z^\star_1(z_0)\in \mathbb Z^\star_{J_0}$) as long as $\ell(z_0)$ is not sufficiently small. Consequently all the arguments above can be repeated with $z^\star_1(z_0)$ playing the role of the new initial extended state in order to prove that \eqref{decrease} holds for all $k$ such that $\ell(z^{\text{cl}}_k)$ is not sufficiently small. But this is precisely \eqref{asymptotic}. $\hfill\Box$
\section{Illustrative example}\label{secillustrative}
Let us consider the $6$-states $2$-control Planar Vertical Take-of and Landing aircraft governed by the following normalized ODEs:
\begin{align}
\ddot y_1&=-u_1\sin\theta+\mu u_2\cos\theta\\
\ddot y_2&=+u_1\cos\theta+\mu u_2\sin\theta-1\\
\ddot\theta&=u_2
\end{align}
where $\mu=0.4$ is a fixed parameter. $y_1$, $y_2$, $\theta$ are the planar coordinates and the roll angle to be controlled around some desired set-point. Denoting by $x=(y_1,y_2,\theta,\dot y_1, \dot y_2, \dot\theta)\in \mathbb R^6$ the state vector. The set-point is denoted by $x^{\text{ref}}=(y_1^{\text{ref}},y_2^{\text{ref}},0,0,0,0)$. The implementation uses a sampling period of $\tau=0.1$ and a prediction horizon $N=15$. Consider the following stage cost expressing this objective:
\begin{equation}
\ell(z):=\|x-x^{\text{ref}}\|_Q^2+\|u-u^d\|_R^2 \label{defdeellpvtol}
\end{equation}
where $Q=\texttt{diag}(10^2,10,10,1,1,1)$ and $R=\mathbb I_{2\times 2}$. We also consider a terminal penalty of the form $\|x-x^{\text{ref}}\|_{Q_f}^2$ with $Q_f=100\times Q$ in order to enhance the stability {  of the conventional formulation only}. This defines the standard stage cost and its associated standard penalty to which we might add the  term $\Phi_\gamma$ introduced in the present contribution. This leads to the following cost function ($\delta_{0,\gamma}$ is the Kronecker notation):
\begin{equation}
  \Bigl[\gamma^2\|f_c(x_N,u_N)\|+\gamma\ell(x_N)\Bigr]+\underbrace{\texttt{standard cost}}_{\delta_{0,\gamma}\cdot\Psi_f(x_N)+\sum_{i=0}^N \ell(x_i,u_i)} \label{}
\end{equation}
In this section, three settings of the above cost function are compared: 
\begin{itemize}
\item[1.] The nominal version with $\gamma=0$.
\item[2.] The proposed version with $\gamma>0$ as proposed in the present paper. 
\item[3.] The version with $\gamma>0$ that does not include the first term involving $\gamma^2$, namely only the $\gamma\ell(x_N)$ is used. 
\end{itemize}
when $\gamma>0$ (options 2. and 3.) several value of $\gamma\in \{1,5,50,100,1000,5000\}$ are tested. \e The following constraints are considered in the implementation: 
\begin{equation}
u\in [-1.5,+1.5]\times [-0.5,+0.5]\ ,\  \vert \dot z\vert\le 0.3\ ,\ \  \vert \dot\theta\vert < 0.2\label{satconstr}
\end{equation}
Starting from $x_0=0$ four different target states are considered, namely:
\begin{equation}
x^{\text{ref}}\in \Bigl\{(d,d,0,0,0,0)\quad\vert\quad d\in \{0.2, 0.5, 1.0, 2.0\}\Bigr\} \label{defdexref}
\end{equation}
where, {  because of the saturation constraints \eqref{satconstr} on the derivatives, only the first target ($d=0.2$) corresponds to the standard $N$-reachability condition being satisfied, the other values leads to this condition being increasingly violated. Indeed the maximum rate of variation of $z$ is 0.3 meaning that the maximum increment over a prediction horizon of $N\tau=1.5$ is upper bounded by $0.45$}. The NMPC has been implemented using \texttt{python-Casadi} framework with the \texttt{IPOPT} solver limited to \texttt{Max\_iter=15} number of iteration in order to meet real-time implementability condition [see Figure \ref{fig2}]. 
\begin{figure*}
\begin{center}
\textbf{(a) With penalty on final derivative activated}.\e 
($d=0.2$) \hskip 2.5cm ($d=0.5$) \hskip 2.8cm ($d=1.0$) \hskip 2.6cm ($d=2.0$)
\e 
\includegraphics[width=0.9\textwidth]{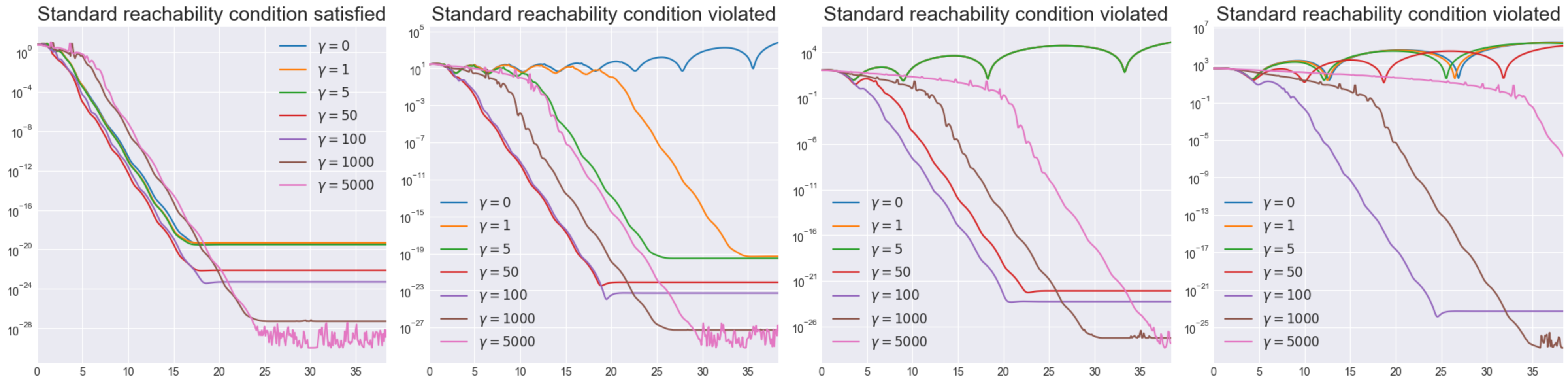} \\
\textbf{(b) Without penalty on the final derivative}.\e 
\includegraphics[width=0.9\textwidth]{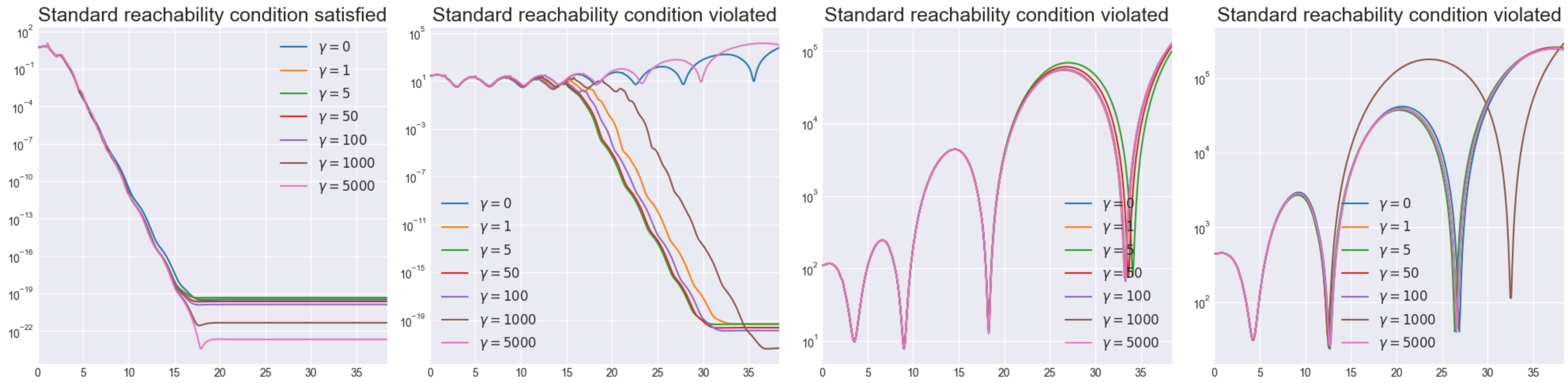} 
\end{center}
\caption{  Closed-loop evolution of the open-loop optimal cost $J^\star(z_k^\text{cl})$ under the different settings and targeted states.}\label{fig1}
\end{figure*}
\begin{figure*}
\begin{center}
\includegraphics[width=0.5\textwidth]{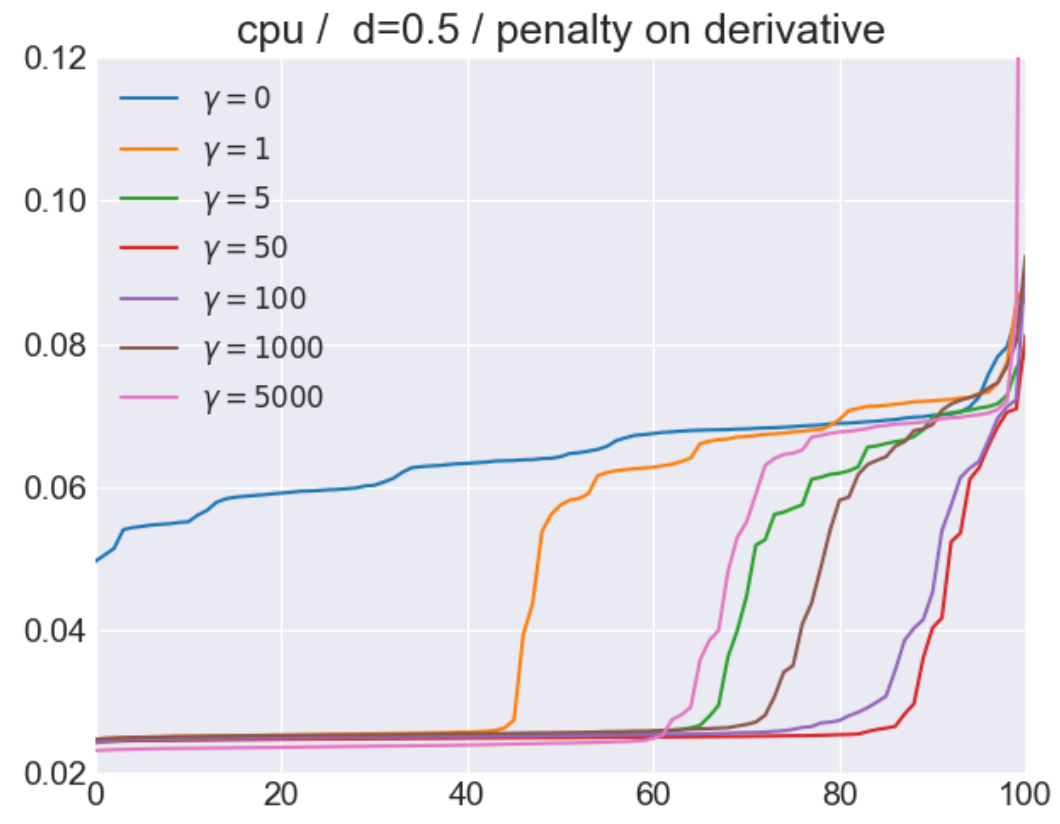} 
\caption{Histogram of \texttt{cpu}'s (\texttt{Apple MacBook Pro, M3, 18Go}).}\label{fig2}
\end{center}
\end{figure*}
Figure \ref{fig1} shows that when the standard $N$-reachability condition is satisfied $(d=0.2)$, all settings lead to successful behaviour. When this condition is strongly violated ($d\ge 1$), only the proposed formulation with sufficiently high $\gamma$ are successful (see the last plots on Figure \ref{fig1}.a). In particular, it is useless to increase $\gamma$ while no penalty on the final derivative is applied as suggested by the last two plots of Figure \ref{fig1}.b. When standard reachability condition is almost satisfied ($d=0.5$), the setting without the derivative ($\gamma^2$)-related penalty might be successful provided that $\gamma$ is not too high. Finally Figure \ref{fig2} shows that the use of high values of $\gamma$ (except for $\gamma=5000$) does not have detrimental effect on the computation time if not the contrary. 
\section{Conclusion}
In this paper, a new provably stable formulation is proposed for NMPC that is appropriate to situations where the standard $N$-reachability condition is not satisfied due to the use of too short prediction horizons. The associated sufficient conditions are discussed and a proof for the practical stability is provided. The results highlighted the relevance of the terminal penalty on the norm of the derivative of the state vector in inducing stability and suggests its systematic use in NMPC formulations. 
\bibliography{bibliography}

\begin{thebibliography}{11}
\expandafter\ifx\csname natexlab\endcsname\relax\def\natexlab#1{#1}\fi
\providecommand{\url}[1]{\texttt{#1}}
\providecommand{\href}[2]{#2}
\providecommand{\path}[1]{#1}
\providecommand{\DOIprefix}{doi:}
\providecommand{\ArXivprefix}{arXiv:}
\providecommand{\URLprefix}{URL: }
\providecommand{\Pubmedprefix}{pmid:}
\providecommand{\doi}[1]{\href{http://dx.doi.org/#1}{\path{#1}}}
\providecommand{\Pubmed}[1]{\href{pmid:#1}{\path{#1}}}
\providecommand{\bibinfo}[2]{#2}
\ifx\xfnm\relax \def\xfnm[#1]{\unskip,\space#1}\fi
%Type = Article
\bibitem[{Alamir(2017)}]{alamir2017contraction}
\bibinfo{author}{Alamir, M.}, \bibinfo{year}{2017}.
\newblock \bibinfo{title}{Contraction-based nonlinear model predictive control formulation without stability-related terminal constraints}.
\newblock \bibinfo{journal}{Automatica} \bibinfo{volume}{75}, \bibinfo{pages}{288--292}.
%Type = Article
\bibitem[{Alamir(2024)}]{alamir2024derivation}
\bibinfo{author}{Alamir, M.}, \bibinfo{year}{2024}.
\newblock \bibinfo{title}{Derivation of certification-based admissibility dashboard of {NMPC} implementation settings: Framework and associated python package}.
\newblock \bibinfo{journal}{IEEE Transactions on Control Systems Technology} .
%Type = Article
\bibitem[{Alamir and Pannocchia(2021)}]{alamir2021new}
\bibinfo{author}{Alamir, M.}, \bibinfo{author}{Pannocchia, G.}, \bibinfo{year}{2021}.
\newblock \bibinfo{title}{A new formulation of economic model predictive control without terminal constraint}.
\newblock \bibinfo{journal}{Automatica} \bibinfo{volume}{125}, \bibinfo{pages}{109420}.
%Type = Article
\bibitem[{Garone et~al.(2017)Garone, Di~Cairano and Kolmanovsky}]{garone2017reference}
\bibinfo{author}{Garone, E.}, \bibinfo{author}{Di~Cairano, S.}, \bibinfo{author}{Kolmanovsky, I.}, \bibinfo{year}{2017}.
\newblock \bibinfo{title}{Reference and command governors for systems with constraints: A survey on theory and applications}.
\newblock \bibinfo{journal}{Automatica} \bibinfo{volume}{75}, \bibinfo{pages}{306--328}.
%Type = Article
\bibitem[{Klau{\v{c}}o et~al.(2017)Klau{\v{c}}o, Kaluz and Kvasnica}]{klauvco2017real}
\bibinfo{author}{Klau{\v{c}}o, M.}, \bibinfo{author}{Kaluz, M.}, \bibinfo{author}{Kvasnica, M.}, \bibinfo{year}{2017}.
\newblock \bibinfo{title}{Real-time implementation of an explicit mpc-based reference governor for control of a magnetic levitation system}.
\newblock \bibinfo{journal}{Control Engineering Practice} \bibinfo{volume}{60}, \bibinfo{pages}{99--105}.
%Type = Inproceedings
\bibitem[{K{\"o}hler et~al.(2018)K{\"o}hler, M{\"o}ller and Allg{\"o}wer}]{kohler2018nonlinear}
\bibinfo{author}{K{\"o}hler, J.}, \bibinfo{author}{M{\"o}ller, M.A.}, \bibinfo{author}{Allg{\"o}wer, F.}, \bibinfo{year}{2018}.
\newblock \bibinfo{title}{Nonlinear reference tracking with model predictive control: An intuitive approach}, in: \bibinfo{booktitle}{2018 European Control Conference (ECC)}, \bibinfo{organization}{IEEE}. pp. \bibinfo{pages}{1355--1360}.
%Type = Techreport
\bibitem[{La~Salle(1966)}]{la1966invariance}
\bibinfo{author}{La~Salle, J.P.}, \bibinfo{year}{1966}.
\newblock \bibinfo{title}{An invariance principle in the theory of stability}.
\newblock \bibinfo{type}{Technical Report}.
%Type = Article
\bibitem[{Mayne et~al.(2000)Mayne, Rawlings, Rao and Scokaert}]{Mayne2000}
\bibinfo{author}{Mayne, D.Q.}, \bibinfo{author}{Rawlings, J.}, \bibinfo{author}{Rao, C.V.}, \bibinfo{author}{Scokaert, P.O.M.}, \bibinfo{year}{2000}.
\newblock \bibinfo{title}{Constrained model predictive control: Stability and optimality}.
\newblock \bibinfo{journal}{Automatica} \bibinfo{volume}{36}, \bibinfo{pages}{789--814}.
%Type = Article
\bibitem[{M\"{u}ller et~al.(2014)M\"{u}ller, Angeli, Allg\"{o}wer, Rish and Rawlings}]{Muller2014}
\bibinfo{author}{M\"{u}ller, M.A.}, \bibinfo{author}{Angeli, D.}, \bibinfo{author}{Allg\"{o}wer, F.}, \bibinfo{author}{Rish, A.}, \bibinfo{author}{Rawlings, J.B.}, \bibinfo{year}{2014}.
\newblock \bibinfo{title}{Convergence in economic model predictive control with average constraints}.
\newblock \bibinfo{journal}{Automatica} \bibinfo{volume}{50}, \bibinfo{pages}{3100 -- 3111}.
%Type = Article
\bibitem[{Polver et~al.(2025)Polver, Limon, Previdi and Ferramosca}]{polver2025robust}
\bibinfo{author}{Polver, M.}, \bibinfo{author}{Limon, D.}, \bibinfo{author}{Previdi, F.}, \bibinfo{author}{Ferramosca, A.}, \bibinfo{year}{2025}.
\newblock \bibinfo{title}{Robust contraction-based model predictive control for nonlinear systems}.
\newblock \bibinfo{journal}{arXiv preprint arXiv:2502.02394} .
%Type = Book
\bibitem[{Rawlings et~al.(2017)Rawlings, Mayne, Diehl et~al.}]{rawlings2017model}
\bibinfo{author}{Rawlings, J.B.}, \bibinfo{author}{Mayne, D.Q.}, \bibinfo{author}{Diehl, M.}, et~al., \bibinfo{year}{2017}.
\newblock \bibinfo{title}{Model predictive control: theory, computation, and design}. volume~\bibinfo{volume}{2}.
\newblock \bibinfo{publisher}{Nob Hill Publishing Madison, WI}.

\end{thebibliography}
\bibliographystyle{elsarticle-harv}
\end{document}